\documentclass[letterpaper, 10 pt, conference]{ieeeconf}  

\IEEEoverridecommandlockouts                              

\usepackage{cite}
\usepackage{subfig}
\usepackage{amsmath,amsthm,amssymb,amsfonts}
\usepackage{graphicx}
\usepackage[colorinlistoftodos]{todonotes}
\usepackage{booktabs}
\usepackage{array}
\let\labelindent\relax
\usepackage[shortlabels]{enumitem}
\newtheorem{theorem}{Theorem}

\theoremstyle{definition}
\newtheorem{definition}{Definition}
\usepackage{algorithm}
\usepackage{algpseudocode}

\newcolumntype{F}{>{\centering\arraybackslash}m{20em}}

\usepackage{tikz}
\usepackage{textcomp}
\usepackage{lipsum}

\renewcommand{\baselinestretch}{1}

\newcommand{\copyrighttext}{%
  \footnotesize \textcopyright 20XX IEEE.  Personal use of this material is permitted.  Permission from IEEE must be obtained for all other uses, in any current or future media, including reprinting/republishing this material for advertising or promotional purposes, creating new collective works, for resale or redistribution to servers or lists, or reuse of any copyrighted component of this work in other works. 
DOI: 
}
\newcommand{\copyrightnotice}{%
\begin{tikzpicture}[remember picture,overlay]
\node[anchor=south,yshift=10pt] at (current page.south) {\fbox{\parbox{\dimexpr\textwidth-\fboxsep-\fboxrule\relax}{\copyrighttext}}};
\end{tikzpicture}%
}

\title{\LARGE \bf
Correct-By-Construction Design of Adaptive Cruise Control with Control Barrier Functions Under Safety and Regulatory Constraints}

\author{Muhammad Waqas$^{1}$, Muhammad Ali Murtaza$^{2}$, Pierluigi Nuzzo$^{1} $ and   Petros Ioannou$^{1} $
\thanks{$^{1}$M. Waqas, P. Nuzzo, and P. Ioannou are with the Viterbi School of Engineering, University of Southern California, 90089, USA. Email:
        {\tt\small \{waqas,nuzzo,ioannou\}@usc.edu}.}%
\thanks{$^{2}$M. A. Murtaza is with the Electrical and Computer Engineering Department, Georgia Institute of Technology, Atlanta, USA. Email: {\tt\small mamurtaza@gatech.edu}. 
} %
}
\begin{document}

\maketitle
\hfill
\copyrightnotice
\thispagestyle{empty}
\pagestyle{empty}

\begin{abstract}
The safety-critical nature of adaptive cruise control (ACC) systems calls for systematic design procedures, e.g., based on formal methods or control barrier functions (CBFs), to provide strong guarantees of safety and performance under all driving conditions. However, existing approaches have mostly focused on fully verified solutions under smooth traffic conditions, with the exception of stop-and-go scenarios. Systematic methods for high-performance ACC design under safety and regulatory constraints like traffic signals are still elusive. A challenge for correct-by-construction approaches based on CBFs stems from the need to capture the constraints imposed by traffic signals, which lead to candidate time-varying CBFs (TV-CBFs) with finite jump discontinuities in bounded time intervals.
%
This paper addresses this challenge 
by showing how traffic signal constraints can be effectively captured in the form of piecewise continuously differentiable TV-CBFs, from which we can generate 
 switching-based controllers that are guaranteed to be safe and comply with regulatory signals. 
Simulation results show the effectiveness of the proposed approach.

\end{abstract}


\section{Introduction}

The goal of adaptive cruise control (ACC)~\cite{ioannou1993intelligent,marsden2001towards} is to ensure that the vehicle under control, i.e., the \emph{ego} vehicle, tracks the velocity of the leading vehicle while maintaining a safe distance. The safe distance is usually calculated by using a constant-time headway policy, the headway time being 
the time the ego vehicle takes to cover the 
distance between itself and the leading vehicle. 
ACC systems have been extensively studied over the last decade. Predictive cruise control~\cite{asadi2010predictive} uses time sequence information from upcoming traffic signals to optimize fuel efficiency for vehicle planning. Similarly, ecological ACC~\cite{bae2019design} aims to avoid traffic signal violations and collisions while generating optimal reference velocity signals to minimize fuel consumption. These optimization-based approaches, however, tend to lack strong guarantees that the ego vehicle is safe and obeys regulatory constraints. 
More recently, the safety-critical nature of ACC systems has called for formally verified or correct-by-construction approaches using methods from theorem proving~\cite{loos2011adaptive}, algorithmic control synthesis~\cite{nilsson2015correct}, and control barrier functions (CBFs)~\cite{mehra2015adaptive,ames2014control,xu2017correctness} to provide strong guarantees of safety and performance. 

State-of-the-art formal verification and correct-by-construction design methods have been successfully applied in the context of highway systems with smooth traffic conditions. Recently,  a provably correct ACC design approach has been proposed to safely handle the occurrence of cut-in vehicles while preserving comfort in a model predictive control (MPC) scheme~\cite{althoff2020provably}. 
However, control synthesis methods that can deal with regulatory constraints like non-smooth traffic signals are still elusive. In this paper, we focus on the synthesis of adaptive cruise controllers under safety and regulatory constraints like traffic signals, a class of systems that we call \emph{regulated ACCs}, using control barrier guarantees. 

We model a traffic signal as a function of time, e.g., $s: [0,\infty) \rightarrow \{ \mathtt{Green}, \mathtt{Yellow}, \mathtt{Red}\}$, that exhibits finite jump discontinuities within bounded time intervals. Capturing the traffic signal constraints in the form of CBFs leads to time-varying CBFs (TV-CBFs) with jump discontinuities, which makes it difficult to apply standard CBF-based design methods. In fact, non-smooth barrier functions (NBFs)~\cite{glotfelter2019hybrid} have been investigated for time-invariant CBFs. In the  time-varying case,  multiple CBFs can be combined via a pointwise minimum operator~\cite{lindemann2018control}.
However, applying this method to traffic light signals, for example, would require that the vehicle stop at the stop line of every traffic signal, be it green or red, which is overly conservative for practical scenarios, as shown with examples in Section~\ref{Sec:Preliminaries}. 
%
%
We propose, instead, to \emph{represent traffic signals with jump discontinuities via piecewise $m$-times continuously differentiable ($\mathcal{C}^m$) TV-CBFs and investigate conditions for the existence of 
switching-based controllers that render the corresponding safe sets forward-invariant}. 
Our contributions can be summarized as follows: 

\begin{itemize}
    \item We present a control synthesis method for piecewise $m$-times continuously differentiable ($\mathcal{C}^m$) TV-CBFs with finite jump discontinuities within bounded time intervals. We prove that the super-level set of such a 
    TV-CBF is forward-invariant under a switching-based controller.  
    \item Based on the method above, we design a correct-by-construction regulated ACC, which  receives the traffic lights' time sequence and guarantees it will obey these signals while keeping safe spacing with leading vehicles and limiting the velocity of the ego vehicle to a maximum value set by the driver. 
\end{itemize}


We organize the paper as follows. We provide an overview of CBF-based methods in Section~\ref{Sec:Preliminaries}. We then introduce the piecewise $\mathcal{C}^m$ TV-CBFs in Section~\ref{Sec:CBF} and formulate the regulated ACC design problem in Section~\ref{Sec: Regulated-ACC}. In Section \ref{Sec:CBF-ACC}, a piecewise $\mathcal{C}^m$ TV-CBF is constructed for the regulated-ACC problem. In Section \ref{Sec:Controller Design}, the controller is synthesized from the CBF constraints via quadratic programming. Simulation results and conclusions are presented in Section~\ref{Sec:Simulation} and  Section~\ref{Sec:Conclusion}, respectively.

\section{Preliminaries\label{Sec:Preliminaries}}

In this section, we provide an overview of control barrier functions (CBFs) and motivate our design problem in this context. In the following, 
$\mathcal{C}^m$ denotes the class of $m$-times continuously differentiable functions defined on $D\subset \mathbb{R}^n$. A continuous function $\alpha:[0,a)\rightarrow \mathbb{R}_{\geq 0}$ is a class-$\mathcal{K}$ function when $\alpha(0)=0$ and $\alpha$ is strictly monotonically increasing. 
{Given $h:\mathbb{R}^n \rightarrow \mathbb{R}$ and $f:\mathbb{R}^n \rightarrow \mathbb{R}^n$, the Lie derivative of $h(\mathbf{x})$ with respect to $f(\mathbf{x})$ is defined as $L_fh(\mathbf{x})=\frac{dh(\mathbf{x})}{d\mathbf{x}}f(\mathbf{x})$. Given $h:[0,\infty) \times \mathbb{R}^n \rightarrow \mathbb{R}$, the Lie derivative is defined as $L_fh(t,\mathbf{x})=\frac{\partial h(t,\mathbf{x})}{\partial \mathbf{x}}f(\mathbf{x})$.} Finally, a function $f: D \rightarrow \mathbb{R}^n$ is said to be locally Lipschitz on its domain $D$ if $\forall x\in D$, $\exists$ a neighbourhood $D_0 \subseteq D$ such that $\forall x,y\in D_0$, $\exists\;L$ such that $\parallel f(x)-f(y) \parallel \leq L \parallel x-y \parallel$.

We start by considering the following control-affine system:
\begin{equation}
\label{eq:dx=f+gu}
\dot{\mathbf{x}} = f(\mathbf{x}) + g(\mathbf{x})u,
\end{equation}
where $\mathbf{x}\in \mathbb{R}^n$ is the state, $u\in \mathcal{U} \subseteq \mathbb{R}^q$ is the control input, $\mathcal{U}$ being the set of allowed inputs, and $f:\mathbb{R}^n\rightarrow \mathbb{R}^n$ and $g:\mathbb{R}^n\rightarrow \mathbb{R}^q$ are locally Lipschitz functions. 
We would like to design a controller that guarantees the safety of system~(\ref{eq:dx=f+gu}), where the safe set $C$ is defined as the superlevel set of a continously differentiable function $h(\mathbf{x}):D\rightarrow \mathbb{R}$. We define $C$, its boundary $\partial C$, and its interior  $\mathrm{Int}(C)$ as follows:
\begin{align*}
\begin{split}
C=\{\mathbf{x}\in \mathbb{R}^n:h(\mathbf{x})\geq 0\},\\
\partial C=\{\mathbf{x}\in \mathbb{R}^n:h(\mathbf{x})= 0\},\\
\mathrm{Int} C =\{\mathbf{x}\in \mathbb{R}^n:h(\mathbf{x})> 0\}.
\end{split}
\end{align*}
Then, if $h(\mathbf{x})$ is a CBF, such a controller is guaranteed to exist. To formally state this result, we recall the notions of forward-invariant set and CBF.

\begin{definition}[Forward-Invariant Set~\cite{ames2019control}]
$C$ is said to be forward-invariant for system~(\ref{eq:dx=f+gu}) when, $\forall \ \mathbf{x}(t_0) \in C$, $\exists \ u=k(\mathbf{x})$ such that $\mathbf{x}(t)\in C$ for all $t\geq t_0$. In other words, if the system state is initially in $C$, then there exists a controller that ensures that the system always stays in $C$. 
\end{definition}

\begin{definition}[Control Barrier Function~\cite{ames2019control}] \label{def:CBF}
Let $h(\mathbf{x}):D \rightarrow \mathbb{R}$ be a continuously differentiable function and $C$ be the corresponding superlevel set. Furthermore, let $\frac{dh}{d\mathbf{x}}\neq 0$ $\forall\; \mathbf{x}\in \partial C$. 
If there exists a class-$\mathcal{K}$ function $\alpha$ such that, for the system in~(\ref{eq:dx=f+gu}), $\forall\; \mathbf{x}\in D$, the following holds 
\begin{equation*}
\sup\limits_{u\in \mathcal{U}}[L_{f}h(\mathbf{x})+L_{g}h(\mathbf{x})u]\geq-\alpha(h(\mathbf{x})),
\end{equation*}
then $h(\mathbf{x})$ is called a control barrier function. 
\end{definition}

We can then state the main result of this section, characterizing the set of safe inputs which make system~\ref{eq:dx=f+gu} safe with  respect to the superlevel set $C$ of a CBF $h(\mathbf{x})$. 

\begin{definition}[Set of Safe Inputs~\cite{ames2019control}] \label{def:safeset}
The set of safe inputs that renders the superlevel set $C$ of $h(\mathbf{x})$ forward-invariant for system~\ref{eq:dx=f+gu} is given by $\mathcal{U}_{h}(\mathbf{x})=\{u\in \mathcal{U}:L_{f}h(\mathbf{x})+L_{g}h(\mathbf{x})u+\alpha(h(\mathbf{x}))\geq 0\}$.  Any Lipschitz continous control law of the form $u=k(\mathbf{x})\in \mathcal{U}_{h}$ will render the system safe. \end{definition}

\subsection{Time-Varying Control Barrier Functions}

The result in Definition~\ref{def:safeset} relates to a time-invariant CBFs. A similar result can, however, be stated for time-varying CBFs~\cite{xiao2019control}. We first recall the notion of relative degree of a CBF, since higher-order CBFs are often required to express many constraints in motion planning. 

\begin{definition}[Relative Degree~\cite{ames2019control,xiao2019control}]
A time-invariant CBF $h(\mathbf{x})$ has relative degree $r$ when $L_gL_f^{r-1}h(\mathbf{x})\neq 0$ and $L_gL_f^ih(\mathbf{x})=0$ for $i=0,1,2,...,r-2$. A time-varying CBF has relative degree $r$ when $L_gL_f^{r-1}h(t,\mathbf{x})\neq 0$ and $L_gL_f^ih(t,\mathbf{x})=0$ for $i=0,1,2,...,r-2$.
\end{definition} 

If $h(t,\mathbf{x})$ is a time-varying higher-order CBF (HOCBF)~\cite{xiao2019control} in $\mathcal{C}^m$, i.e., a CBF with relative degree $m > 1$, then there exist safe controllers that guarantee the forward-invariance of a time-varying set $C(t)$ defined as follows. 

\begin{definition}[Time-Varying Higher Order CBF\cite{xiao2019control}]\label{def:HOCBF}
Let $h(t,\mathbf{x}): \Gamma\times \mathcal{D} \rightarrow \mathbb{R}$ be an $m$-times continuously differentiable function and let  $\beta_i(t,\mathbf{x})$, $\forall i \in \{0,1,2, \ldots,m\}$, be defined as follows: 
\begin{equation}
\label{HOCBF}
\begin{array}{c} {{\beta _0}(t,{\mathbf{x}}): = h(t,{\mathbf{x}})}, \\ {{\beta _1}(t,{\mathbf{x}}): = {{\dot \beta }_0}({\mathbf{x}},t) + {\alpha _1}\left( {{\beta _0}(t,{\mathbf{x}})} \right),} \\ \vdots \\ {{\beta _m}(t,{\mathbf{x}}): = {{\dot \beta }_{m - 1}}(t,{\mathbf{x}}) + {\alpha _m}\left( {{\beta _{m - 1}}(t,{\mathbf{x}})} \right)}, \end{array}
\end{equation}
where $\Gamma\in \{(t_0,t_1), [t_0,t_1), [t_0,t_1],(t_0,t_1]\}$ and the $\alpha_i$ are class-$\mathcal{K}$ functions. We define the super-level sets $C_i(t)=\{\mathbf{x}\in \mathbb{R}^n: \beta_{i-1}(t,\mathbf{x}) \geq 0\}$, for $ i=\{1,2,3, \ldots, m\}$. 

The function $h(t,\mathbf{x})$ is called time-varying HOCBF if there exist $m$ continuously differentiable class-$\mathcal{K}$ functions $\alpha_1,\alpha_2,\dots, \alpha_m$ such that the following holds:
\begin{gather} 
\label{eq:u for HOCBF}
\begin{split}
L_f^mh({\mathbf{x}},t) + {L_g}L_f^{m - 1}h({\mathbf{x}},t){u} + \frac{{{\partial ^m}h({\mathbf{x}},t)}}{{\partial {t^m}}} \\ + O(h({\mathbf{x}},t)) + {\alpha _m}\left( {{\beta _{m - 1}}({\mathbf{x}},t)} \right) \geq 0, 
\end{split}
\end{gather}
where $O(\cdot)$ denotes the remaining Lie derivatives in the direction of $f$ and the partial derivatives from degree $1$ to degree $m-1$. Furthermore, any $u\in \mathcal{U}$ that
satisfies~(\ref{eq:u for HOCBF}) renders the set $C(t)=\underset{1 \leq i \leq m}\bigcap C_i(t)$ forward-invariant for system~(\ref{eq:dx=f+gu}). 
\end{definition}

In other words, if a TV-CBF  $h(t,\mathbf{x}) \in \mathcal{C}^m$ with $m\geq 1$, then any controller that satisfies~(\ref{eq:u for HOCBF}) is safe with respect to $C(t)$. 
However, in many robotic and vehicular planning applications, the safety set of a system varies with time in a discontinuous manner. Such safety sets could naturally be represented in terms of piecewise $\mathcal{C}^m$ CBFs. However, in the presence of discontinuities, the above results from HOCBFs cannot be directly used. To be able to use the properties of the  TV-CBFs in Definition~\ref{def:HOCBF},  
we may try to design a $\mathcal{C}^m$ TV-CBF that can capture a smooth approximation of $C(t)$. However, this can lead to overly conservative designs, as further 
elaborated using the following example.   

\subsection{Incorporating Traffic Rules Within  CBFs\label{subsec:CBF-Traffic-Signal}}

Let us consider a vehicle with simple dynamics $\dot{x}=u$, where $x\in \mathbb{R}$ and $u\in \mathbb{R}_{\geq 0}$ denote the position of the vehicle and its velocity, respectively. Let $\mathbf{x}=\begin{bmatrix} x &v\end{bmatrix}^T$ be the state of the system. Let there be $n$ traffic signals, each at position $p_i$, $i \in \{1, \ldots, n\}$. The current state of the $i^{th}$ traffic signal is denoted by $s_i\in\{\mathtt{Green},\mathtt{Yellow},\mathtt{Red}\}$ and its time sequence 
is given by $(g_{i1}, y_{i1},r_{i1},g_{i2},y_{i2},r_{i2},g_{i3},y_{i3},r_{i3}, \ldots)$. In this sequence, $g_{i1}$ denotes the time at which the signal turns $\mathtt{Green}$ for the first time and, similarly, $g_{ij}$, $y_{ij}$, and $r_{ij}$ denote the $j^{th}$ occurrence of a transition of the $i^{th}$ signal to $\mathtt{Green}$, $\mathtt{Yellow}$, and $\mathtt{Red}$, respectively. 

\begin{figure}[t]
\includegraphics[width=\columnwidth]{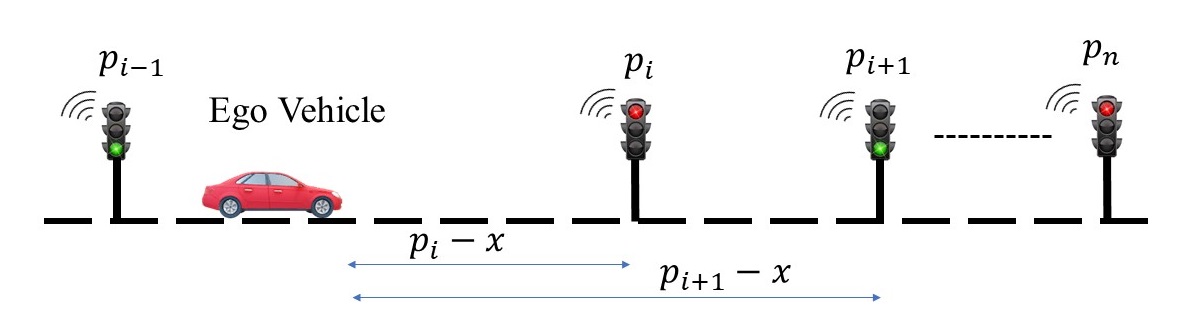}
\caption{Ego vehicle approaching the $i^{th}$ traffic signal at $p_i$.}
\label{fig:N_signals}
\end{figure}

Let the vehicle approach a traffic signal at $p_i$ and suppose that $p_{i-1}<x\leq p_i$ holds, as shown in Fig.~\ref{fig:N_signals}. 
The vehicle should not go beyond the stop line at  $p_i$ when the $i^{th}$ traffic signal is red. When the $i^{th}$ traffic signal is green or yellow, then the vehicle can cross $p_i$. Therefore, when the ego vehicle is between $p_{i-1}$ and $p_i$, the safe set can be expressed as
\begin{align}
\label{eq:safe_set_of_non_valid_CBF}
\mathfrak{C}(t) &=\begin{cases}
\mathbf{x}\in \mathbb{R}^2: p_{i+1}- x\geq 0& \mbox{if }s_i \neq \mathtt{Red},
\\
\mathbf{x}\in \mathbb{R}^2: p_{i}- x\geq 0& \mbox{if }s_i= \mathtt{Red},
\end{cases} 
\end{align}
suggesting the following candidate CBF
\begin{align}
\label{eq:non_valid_CBF}
\begin{split}
\mathfrak{h}_i(t,\mathbf{x})&=\begin{cases}
p_{i+1}-x & g_{ij} \leq t < r_{ij},  j \geq 1,\\
p_{i}-x & r_{ij} \leq t <g_{i,j+1}, j \geq 1. \\
\end{cases}
\end{split}
\end{align}

This candidate TV-CBF is piecewise $\mathcal{C}^1$ in time and it has jump discontinuities at $t = r_{ij}$ and $t = g_{ij}$ for all $j \geq 1$. {Let $\Gamma_{ij}=[g_{ij},g_{i,j+1})$ be the $j^{th}$ cycle of the $i^{th}$ traffic signal. Let $\mathfrak{h}_{ig}(\mathbf{x})=p_{i+1}-x$ be the CBF and $\mathfrak{C}_{ig}=\{\mathbf{x}\in \mathbb{R}^2: p_{i+1}-x \geq 0\}$ be the safe set when $s_i \neq \mathtt{Red}$. Let $\mathfrak{h}_{ir}(\mathbf{x})=p_{i}-x$ be the CBF and $\mathfrak{C}_{ir}=\{\mathbf{x}\in \mathbb{R}^2: p_{i}-x \geq 0\}$ be the safe set when $s_i=\mathtt{Red}$. 
A sufficient condition to ensure safety over $\Gamma_{ij}$ is to require that the minimum of the two safe sets be forward-invariant, which can be achieved by selecting the pointwise minimum between $\mathfrak{h}_{ig}$ and $\mathfrak{h}_{ir}$. By considering that $\mathfrak{C}_{ir} \subset \mathfrak{C}_{ig}$ and by taking a smooth under-approximation of the pointwise minimum~\cite{lindemann2018control} over all  
cycles, $\forall t\in\Gamma_{ij}$, $\forall j\geq 1$, we obtain:}
\begin{align} 
\label{eq:under_approximation_min}
\begin{split}
-\ln \left ({\sum _{j=1, t\in \Gamma_{ij}}^{\infty} \exp (-\mathfrak{h}_{i}(t,\mathbf{x}))}\right) & \le \min _{j \geq 1, t\in \Gamma_{ij}} \mathfrak{h}_{i}(t,\mathbf{x})\\
& \leq  p_{i}-x
\end{split}
\end{align}
%
However, \eqref{eq:under_approximation_min} leads to a very conservative design, as 
it requires that the vehicle stop at the stop line of the upcoming traffic signal, be it $\mathtt{Red}$, $\mathtt{Green}$ or $\mathtt{Yellow}$. 
In the following, we propose a novel method to overcome this issue by directly dealing with piecewise $C^m$ TV-CBFs with finite jump discontinuities in bounded time intervals. 

\section{Piecewise $\mathcal{C}^m$ Control Barrier Functions}\label{Sec:CBF}

We first introduce the notion of piecewise $C^m$ TV-CBF with finite jump discontinuities in any bounded time interval. 

\begin{definition}[Piecewise $\mathcal{C}^m$ Time-Varying CBF]\label{def:piece-wise TV CBF}
Let $\mathcal{N}$ be a finite or infinite set of time indices and let $h(t,\mathbf{x}):[0,\infty)\times \mathcal{D}\rightarrow \mathbb{R}$ be a piecewise $\mathcal{C}^m$ function defined on non-overlapping intervals of the form $\Gamma_i=[t_{i-1},t_i)$ such that
\begin{align}\label{eq:switched:CBF}
\begin{split}
\ \forall \ t \in& \Gamma_i, \forall \ i \in \mathcal{N};\\
h(t,\mathbf{x}) &= h_i(t,\mathbf{x}), C(t)=C_i(t) \\
C_i(t_i^-) &\subseteq C_{i+1}(t_i), \ \forall \ i\in \mathcal{N};\\
\end{split}
\end{align}
If $\forall i\in\mathcal{N}$ $h_i(t,\mathbf{x})$ is an HOCBF of relative degree $m$ with corresponding superlevel set $C_i(t)=\underset{1 \leq j \leq m}\bigcap C_{ij}(t)$,  then $h(t,\mathbf{x})$ is said to be a valid piecewise $\mathcal{C}^m$ TV-CBF.
\end{definition}
%
In Definition~\ref{def:piece-wise TV CBF},  $h(t,\mathbf{x})$ has jump discontinuities at $t=t_1, t_2, t_3, \dots,t_\mathcal{N}$.  We  assumed that $\Gamma_i=[t_{i-1},t_i)$. However, in general, it is sufficient to require that  $\Gamma_i\in \{[t_{i-1}, t_i],[t_{i-1}, t_i),(t_{i-1}, t_i), (t_{i-1}, t_i]\}$, $\underset{i\in \mathcal{N}}\bigcup \Gamma_i = [0,\infty)$, and $  \Gamma_i \cap \Gamma_j = \emptyset , \mbox{if}\: i \neq j$. 
$C(t)$ denotes the superlevel set of the piecewise $\mathcal{C}^m$ TV-CBF in~(\ref{eq:switched:CBF}). Our goal is to design a controller that renders the safe set $C(t)$ forward-invariant. The following theorem describes such controller.
\begin{theorem}\label{Thm:safe_switching_CBF}
Let $h(t,\mathbf{x})$ be a valid piecewise $\mathcal{C}^m$ TV-CBF as in Definition~\ref{def:piece-wise TV CBF}. If at $t_0=0$, $x(0)\in C_1(0)$, then the following controller will render $C(t)$ forward-invariant:
\begin{equation}
\label{eq: switching based Controller}
u(t) = u_i(t) \in  K_{h,i}(t) \quad  \forall \ t \in \Gamma_i, i \in \mathcal{N},
\end{equation}
where $K_{h,i}(t)$ denotes the set of safe inputs that renders $C_i(t)$ forward-invariant and is given by 
\begin{equation*}
\begin{split}
K_{h,i}(t) &= \{u \in \mathcal{U}:L_f^{m}h_i({\mathbf{x}},t) + {L_g}L_f^{m - 1}h_i(t,{\mathbf{x}}){u} \\
&+\frac{{{\partial ^{m}}h_i(t,{\mathbf{x}})}}{{\partial {t^{m}}}} + O(h_i(t,{\mathbf{x}})) \\
&+ {\alpha _{m}}\left( {{\beta _{i,m - 1}}({\mathbf{x}},t)} \right) \geq 0\}.
\end{split}
\end{equation*}
\end{theorem}
\begin{proof} We proceed by induction. Consider the base case when $x(0)\in C_1(0)$. As $h_1(t,\mathbf{x})$ is an HOCBF, there exists a controller $u_1(t)$ that renders the set $C_1(t)$ forward-invariant in the interval $\Gamma_1$. Therefore, we have  $x(t_1^-)\in C_1(t_1^-)$. As $C_1(t_1^-) \subseteq C_2(t_1)$, we conclude that  $x(t_1^-) \in C_2(t_1)$.

For the induction step, without loss of generality, we can assume that $\Gamma_i= [t_{i-1},t_i)$. We need to prove that, if $x(t_{i-1})\in C_i(t_{i-1})$ and $u_i(t)$ is applied for $t_{i-1}\leq t <t_{i}$, then $x(t_i)\in C_{i+1}(t_i)$ holds. As $h_i(t,\mathbf{x})$ is an HOCBF, there exists a controller $u_i(t)\in K_{h,i}$ for all $t$ such that $t_{i-1} \leq t <t_i$. This implies that  $x(t_i^-)=\lim_{t\to t_i^-} x(t) \in C_i(t_i^-)$. Moreover, $\forall i\in \mathcal{N}$, $C_i(t_i^-)\subseteq C_{i+1}(t_i)$ holds. Therefore, we conclude that $x(t_i)\in C_{i+1}(t_i)$.
\end{proof}

Definition~\ref{def:piece-wise TV CBF} requires that $C_i(t_i^-) \subseteq C_{i+1}(t_i)$ holds $\forall\;i\in \mathcal{N}$. The following theorem gives a sufficient condition for this to happen.

\begin{theorem}\label{Thm:safe_on_jump}
Let $h_i(t,\mathbf{x})$ and $h_{i+1}(t,\mathbf{x})$ be two HOCBF of relative degree $m$, as in Definition~\ref{def:HOCBF},  defined in adjacent non-overlapping intervals $\Gamma_i=[t_{i-1},t_i)$ and $\Gamma_{i+1}=[t_i,t_{i+1})$. Let $\{\alpha_1,\alpha_2,\alpha_3,,...,\alpha_m\}$ be $m$-class $\mathcal{K}$ functions for $h_i(t,\mathbf{x})$ and $h_{i+1}(t,\mathbf{x})$. 
If $\forall \: k\in \{0,1,2,...,m-1\}$ we have 
\begin{equation}
\label{eq:safe_composition_CBF}
\frac{d^{k}}{dt^k}h_{i+1}(t_i^+,{\mathbf{x}}) \geq \frac{d^k}{dt^k}h_i(t_i^-,{\mathbf{x}}) 
\end{equation}
then $C_i(t_i^-)\subseteq C_{i+1}(t_i)$ holds.
\end{theorem}
\begin{proof}
We have $\{\alpha_1,\alpha_2,\alpha_3,,...,\alpha_m\}$ be $m$-class $\mathcal{K}$ functions for $h_i(t,\mathbf{x})$ and $h_{i+1}(t,\mathbf{x})$.
{From (\ref{HOCBF}), we have $\beta_{0,i+1}(t_i,\mathbf{x})=h_{i+1}(t_i,\mathbf{x})$ and $h_{0,i}(t_i^-,\mathbf{x})=\beta_{0,i}(t_i^-,\mathbf{x})$. From (\ref{eq:safe_composition_CBF}), we have $h_{i+1}(t_i,\mathbf{x}) \geq h_{0,i}(t_i^-,\mathbf{x})$. Combining these two conditions, we get $\beta_{0,i+1}(t_i,\mathbf{x}) \geq\beta_{0,i}(t_i^-,\mathbf{x})$. From~(\ref{HOCBF}), we also have 
\begin{equation}
\beta_{1,i+1}(t_i,\mathbf{x}) = \dot{\beta}_{0,i+1}(t,\mathbf{x})+ \alpha_1(\beta_{0,i+1}(t,\mathbf{x})).
\end{equation}
Since $\dot{\beta}_{0,i+1}(t,\mathbf{x})= \frac{d}{dt}h_{i+1}(t_i^+,\mathbf{x})\geq \frac{d}{dt}h_i(t_i^-,\mathbf{x})$, we conclude  $\beta_{1,i+1}(t_i,\mathbf{x}) \geq \beta_{1,i}(t_i^-,\mathbf{x})$.}
Similarly, it can be shown that 
\begin{equation*}
\beta_{i+1,k}(t_i,\mathbf{x}) \geq \beta_{i,k}(t_i^-,\mathbf{x}) \geq 0,\;\forall\: k\in \{0,1,2,...,m-1\}.
\end{equation*}
We know that for $h_i(t,\mathbf{x})$,  $C_i(t_i^-)=\underset{1 \leq k \leq m}\bigcap C_{ik}(t_i^-)$, where $C_{ik}(t_i)=\{\mathbf{x}\in \mathbb{R}^n: \beta_{i,k-1}(t_i^-,\mathbf{x}) \geq 0\}$. Similarly, we have $C_{i+1}(t_i)=\underset{1 \leq k \leq m}\bigcap C_{i+1,k}(t_i^-)$, where  $C_{i+1,k}(t_i)=\{\mathbf{x}\in \mathbb{R}^n: \beta_{i+1,k-1}(t_i^-,\mathbf{x}) \geq 0\}$. Therefore, we conclude that $C_i(t_i^-)\subseteq C_{i+1}(t_i)$ holds. 
\end{proof}


\noindent \textbf{Example.} For the system $\dot{x}=u$, let us consider the piecewise $\mathcal{C}^1$ TV-CBF $b(t,\mathbf{x})$. When $t<t_i$, $b(t,\mathbf{x})=b_a(t,\mathbf{x})=100t-x$. When $t\geq t_i$, $b(t,\mathbf{x})=b_b(t,\mathbf{x})=200t-x$.  
This TV-CBF has a jump discontinuity at $t=t_1$. Since we have $b_a(t_1,x) > b_b(t_1^-,x)$, it fulfills the condition in~(\ref{eq:safe_composition_CBF}). Therefore, it is a valid piecewise $\mathcal{C}^1$ TV-CBF, for which a safety controller is guaranteed to exist. 
In the following, we demonstrate that a valid piecewise $\mathcal{C}^m$ TV-CBF can represent traffic constraints.

\subsection{Piecewise $\mathcal{C}^1$ TV-CBF for Traffic Signals}

We consider again the candidate CBF (\ref{eq:non_valid_CBF}) in Section~\ref{subsec:CBF-Traffic-Signal} with corresponding safe set $\mathfrak{C}(t)= \{\mathbf{x}: \mathfrak{h} (t,\mathbf{x}) \geq 0\}$ (\ref{eq:safe_set_of_non_valid_CBF}). 
\begin{align*}
\begin{split}
\mathfrak{h}(t,\mathbf{x}) & = \mathfrak{h}_{i}(t,x) \quad \forall\: g_{i,j} \leq t < g_{i,j+1},\: p_{i-1}< x \leq p_i\\
\mathfrak{h}_i(t,\mathbf{x})&=\begin{cases}
p_{i+1}-x & g_{ij} \leq t < r_{ij},  j \geq 1,\\
p_{i}-x & r_{ij} \leq t <g_{i,j+1}, j \geq 1. \\
\end{cases}
\end{split}
\end{align*}
This function has jump discontinuities at $t=r_{ij}$ and $t=g_{ij}$. At $t=g_{ij}$, we have $ \mathfrak{C}_{i}(g_{i,j+1}^-) \subset \mathfrak{C}_{i}(g_{i,j+1}) $. On the other hand, at $t=r_{ij}$, we observe that $ \mathfrak{C}_{i}(r_{i,j}^-) \not\subseteq \mathfrak{C}_{i}(r_{ij})$. The proposed candidate is not a valid piecewise $\mathcal{C}^1$ TV-CBF in accordance with Definition \ref{def:piece-wise TV CBF}. However, we can modify $\mathfrak{h}(t,\mathbf{x})$ so that it becomes a valid piecewise $\mathcal{C}^m$ TV-CBF such that, at all discontinuities $t_j$, $ \mathfrak{C}(t_j^-) \subset \mathfrak{C}(t_j)$ holds as required by Definition~\ref{def:piece-wise TV CBF}. 

We consider the candidate CBF and safe set defined by the following expressions:
\begin{align}
\begin{split}
\label{eq:ht+pi-xf}
m_{ij}&=\frac{y_{ij}+r_{ij}}{2},\\
h_i(t,\mathbf{x}) &= \underbrace{\left[\frac{p_{i+1} - p_i}{1+e^{\tau(t-m_{ij})}} \right]}_{h_{it}(t)} + \underbrace{p_i -x-S_0}_{h_{ix}(x)},\\
h(t,\mathbf{x}) & = h_{it}(t) + h_{ix}(x),\\ 
&\forall\: g_{i,j} \leq t < g_{i,j+1},\: p_{i-1}< x \leq p_i,\\
C(t)& = \{\mathbf{x}\in \mathbb{R}^2: h(t,\mathbf{x})\geq 0\}, 
\end{split}
\end{align}
where $S_0$ is the safe distance from the stop line of the traffic signal when the \emph{ego} vehicle is completely stopped, $\tau>0$ and $g_{ij} < m_{ij} < {r_{ij}}$ are design parameters. The $\tau$ is the decay rate of $h_{it}(t)$ and $t=m_{ij}$ is the time when $h_{it}(t)$ decays to $\frac{p_{i+1}-p_i}{2}$. For an appropriate choice of $\tau$ and $g_{ij}<m_{ij}<r_{ij}$, we obtain $\mathfrak{h}(t,x)\leq h(t,x)$ and $C(t) \subseteq \mathfrak{C}(t)$. 
For $g_{ij} \leq t< r_{ij}$, $h_{it}(t)$ decays from $p_{i+1}-p_{i}$ to zero. Consequently, $h_{i}(t,\mathbf{x})$ decays smoothly from $h_i(g_{ij},\mathbf{x})\approx p_{i+1}-x-S_0$ to $h_i(r_{ij},\mathbf{x})\approx p_{i}-x-S_0$ as $t$ goes from $g_{ij}$ to $r_{ij}$. Therefore, $h_i(t,\mathbf{x})$ encodes the constraints imposed by traffic signals. The modified $h_i(t,x)$ has a jump discontinuity at $t=g_{ij}$. Moreover, $\forall\; i\in \{1,2,3,\ldots,n\}$ and $\forall\;j\in \mathbb{N}$, the condition in~(\ref{eq:safe_composition_CBF}) is satisfied. At $t=g_{ij}$, we have  $h_i(g_{ij},x)>h_i(g_{ij}^-,x)$. Therefore, $h(t,\mathbf{x})$ is a valid piecewise $\mathcal{C}^1$ TV-CBF, hence the controller in (\ref{eq: switching based Controller}) will render $C(t)$ forward-invariant.

The proposed approach can be extended to the case when the timing sequence of the traffic signals is not known. 
In particular, if we know the minimum time duration of the yellow light of each traffic signal, we can still achieve safety via a more conservative  $h_{it}(t)$. Similarly, if the next traffic signal is too far, we can assume that $s_i=\mathtt{Red}$.
Based on this scenario, the regulated ACC can be formulated in the following manner. 

\section{Regulated Adaptive Cruise Control \label{Sec: Regulated-ACC}}

We start by presenting the longitudinal model for the ego vehicle.

\subsection{Longitudinal Dynamics of Vehicle}

Let the longitudinal position and longitudinal velocity of the ego vehicle and lead vehicle be denoted by $X_f$, $X_l$, and $V_f$, $V_l$ respectively. Let $m$ be the mass of the ego vehicle. We use the following non-linear model for the longitudinal dynamics:

\begin{align}
\begin{split}
\dot{x} &= \frac{d}{dt}\begin{pmatrix} X_f \\V_f \\ X_l  \\ e \end{pmatrix} = \underbrace{ \begin{pmatrix} V_f \\ -\frac{1}{m}F_r \\ V_l  \\ X_r-hV_f-S_0 \end{pmatrix}}_{f(x)} + \underbrace{\begin{pmatrix} 0 \\ \frac{1}{m} \\0 \\0 \end{pmatrix}}_{g(x)}u, \\
F_r &= c_0 + c_1V_f + c_2V_f^2,\\
\end{split}
\end{align}
where the input $u$ is the force applied by the wheels, $c_0$, $c_1$, and $c_2$ are the vehicle parameters, $h>0$ is constant time headway and $F_r$ denotes the sum of all the frictional and aerodynamic forces on the vehicle. The relative distance and the relative velocity of the ego vehicle and lead vehicle are denoted by $X_r=X_l-X_f$ and $V_r=V_l-V_f$. The difference between the actual relative distance and the required relative distance is called the spacing error $\delta$ and it is given by $\delta = X_r - hV_f-S_0$.

\subsection{Problem Formulation}

In ACC systems, the ego vehicle can track the velocity of the lead vehicle by keeping a safe distance. In a regulated ACC, the controller is also subject to regulatory constraints like traffic signals. Specifically, the \emph{regulated} ACC will obey the upcoming traffic signals by utilizing the timing information from them. 

\subsubsection{Assumptions} \label{Assumptions} 

We make the following assumptions:

\begin{enumerate}[(a)]
\item The traffic signals can broadcast the position of the stop line and the timing of the  traffic lights to the environment. The ego vehicle knows when and for how long the traffic lights will turn green, yellow, and red. 
\item The velocities of the vehicles on the road are non-negative.
\item The maximum acceleration limits of the lead vehicle and the ego vehicle are known.
\end{enumerate}
The traffic signals follow the communication protocol below. 

\subsubsection{Protocol of Broadcast}

Each signal broadcasts its timing sequence to its surroundings. Let us suppose that there are $n$ traffic signals and that the stop line associated with the $i^{th}$ traffic signal is located at the position $p_i \in \mathbb{R}_{\geq 0}$. The $i^{th}$ traffic signal broadcasts its timing sequence in the form of a sequence denoted by $S_i= (g_{i1},y_{i1},r_{i1},g_{i2},y_{i2},r_{i2},g_{i3},y_{i3},r_{i3}, \ldots)$, where $i\in \mathbb{N}$. When $g_{ij}\leq t < y_{ij}$ then state $s_i=\mathtt{Green}$. When $y_{ij}\leq t < r_{ij}$ then state $s_i=\mathtt{Yellow}$. Similarly, when $r_{ij}\leq t < g_{i,j+1}$ then state $s_i=\mathtt{Red}$.   

\subsubsection{Specifications}

The specifications are categorized into hard constraints (HCs) and soft constraints (SCs). The regulated ACC must obey the hard constraints at all times. The soft constraints should, instead, be fulfilled as long as the hard constraints are not violated. The constraints of the problem are given below.

\begin{itemize}
\item $\mbox{HC-I}$: The ego vehicle must maintain a relative distance $X_r \geq hV_f+S_0$ or, equivalently, $\delta\geq 0$ must hold $\forall t\geq 0$.
\item $\mbox{HC-II}$: $0\leq V_f \leq V_{max}$, $\forall\:t\geq 0$. 
\item $\mbox{HC-III}$: The ego vehicle must stop at the red signal before the stop line.
\item $\mbox{SC}$: The ego vehicle should track the velocity of the lead vehicle $V_l$ and $\delta$ should go to $0$ as long as the hard constraints $\mbox{HC-I}$, $\mbox{HC-II}$ and $\mbox{HC-III}$ are not violated.
\end{itemize}

In the following section, we formulate the CBF for the regulated ACC.

\section{Control Barrier Function For Regulated ACC}\label{Sec:CBF-ACC}

The CBFs capturing the hard constraints are formulated below.

\subsection{Hard Constraints}

$\mbox{HC-I}$ and $\mbox{HC-II}$ can be represented by following candidate CBFs~\cite{ames2014control,xiao2019control}:
\begin{align}
\begin{split}
h_1(x) &= \delta = X_r-hV_f-S_0,\\
h_2(x) & = V_{max}-V_f.
\end{split}
\end{align}
The corresponding safety sets are given by $C_i= \{x \in \mathbb{R}^4: h_i(x)\geq 0\}$ for $i=1,2$.

Let us denote the TV-CBF corresponding to the traffic signals' constraints by $h_3(t,\mathbf{x})$. Following the encoding in the form of a TV-CBF in~(\ref{eq:ht+pi-xf}), we obtain 
\begin{align}
h_3(t,x)& = h_{3i(t,x)}\;\forall\: g_{i,j} \leq t < g_{i,j+1},\: p_{i-1}< X_f \leq p_i \notag \\
h_{3i}(t,x) &= \underbrace{\left[\frac{p_{i+1} - p_i}{1+e^{\tau(t-m_{ij})}} \right]}_{h_{3it}(t)} + \underbrace{p_i -x_1-S_0}_{h_{3ix}(x)}\\
C_3(t)& = \{x\in \mathbb{R}^4: h_3(t,\mathbf{x})\geq 0\} \notag
\end{align}
The above candidate TV-CBF has relative degree $2$, hence it is a piecewise $\mathcal{C}^2$ function. This function is discontinuous at $t=g_{ij}$, $\forall j\in\mathbb{N}$. At these discontinuities,
condition~(\ref{eq:safe_composition_CBF}) is satisfied: 
\begin{align*}
\begin{split}
h_3(g_{ij},x) &>h_3(g_{ij}^-,x),\\
\frac{dh_3}{dt}(g_{ij},x) &>\frac{dh_3}{dt}(g_{ij}^-,x).  
\end{split}
\end{align*}
By Theorem~\ref{Thm:safe_on_jump}, we conclude  that $C_3(g_{ij}^-)\subset C_3(g_{ij})$, $\forall j\in \mathbb{N}$. Therefore, $h_{3}(t,x)$ is a valid piecewise $\mathcal{C}^2$ TV-CBF. Finally, we fulfill the soft constraints via following nominal controller.

\subsection{Soft Constraints}

We design a nominal PID controller such that $\delta \rightarrow 0$ and $V_f \rightarrow V_l$, as shown below.
\begin{align}
\label{eq:Nominal Controller}
\begin{split}
u_{nom} &= m\mu_{nom}+F_r,\\
\mu_{nom} & = \left(k_1V_r+k_2\delta + k_3 \int_{0}^{t} \delta \,dx\right).
\end{split}
\end{align}
After designing the CBFs, we can synthesize the controller. 

\section{Controller Design\label{Sec:Controller Design}}

We design the controller by using $h_1(\mathbf{x})$, $h_2(\mathbf{x})$, and $h_3(t,\mathbf{x})$ defined above. The set of safe inputs $\mathcal{U}_{h_i}$ which renders the set $C_i$ forward invariant is derived below. Let $\alpha_1(r)=\alpha_2(r) = r$ be the class-$\mathcal {K} $ functions. If $u = F_r+m\mu$, then we get the set of safe inputs as follows:
\begin{align}
\dot{h}_1(x) &= V_r-h\mu, \: \dot{h}_2(x) = -\mu, \notag\\
\mathcal{U}_{h_1}& = \{\mu \in \mathcal{U}:  h\mu \leq V_r + h_1(\mathbf{x})\}, \notag \\
\mathcal{U}_{h_2}& = \{\mu \in \mathcal{U}:  \mu \leq V_{safe}-V_f\}.
\end{align}
\begin{align}
\beta_0(t,\mathbf{x}) &=h_{3i}(t,\mathbf{x}),\: \beta_1(t,\mathbf{x}) =\frac{d \beta_0(t,\mathbf{x})}{d t}+\lambda_1\beta_0(t,\mathbf{x}), \notag \\
\beta_2(t,\mathbf{x}) &=\frac{d \beta_1(t,\mathbf{x})}{dt}+\lambda_2\beta_1(t,\mathbf{x}). 
\end{align}
The values of $\lambda_i$ are selected as follows:
\begin{align}
\begin{split}
\label{eq:lamnda1,2}
\lambda_1 \geq \frac{-\frac{\partial h_{3i}}{\partial t}(x_0,t_0)}{\beta_0(x_0,t_0)}&,\:\:\lambda_2 \geq \frac{-\frac{\partial^2 h_{3i}}{\partial t^2}(x_0,t_0)}{\beta_1(x_0,t_0)}, 
\end{split}
\end{align}
where $t_0=g_{ij}$, $x_0=x(g_{ij})$ and the values of $\lambda_1$ and $\lambda_2$ are chosen such that $\lambda_1>0$ and $\lambda_2>0$. { For $A_{\eta}=\eta(t,\mathbf{x})=\begin{pmatrix} \beta_0(t,\mathbf{x}) & \beta_1(t,\mathbf{x}) \end{pmatrix}^T$, $B_{\eta}=\frac{d}{dt}\eta(t,\mathbf{x}) = \begin{pmatrix} 0 & 1\\ 0 & 0 \end{pmatrix}$, the gain $K_{\alpha}$ can be selected using classical pole-placement methods such that $\dot{\eta}(t,\mathbf{x}) = \left(A_{\eta} - B_{\eta}K_{\alpha}\right)\eta(t,\mathbf{x})$ have the desired poles $\begin{pmatrix} -\lambda_1 & -\lambda_2 \end{pmatrix}$.}  

The set of safe inputs that renders $C_3(t)$ forward-invariant is given as 
\begin{align*}
\begin{split}
\mathcal{U}_{h_3}(t) &= \left\{\mu \in \mathcal{U}: 
\mu \leq \frac{d^2h_{3it}(t)}{dt^2} + K_{\alpha}\begin{pmatrix} h_{3,i}(t,\mathbf{x}) \\  \frac{dh_{3it}}{dt} -x_2 \end{pmatrix}\right\}.
\end{split}
\end{align*}
Combining all the constraints we get $\mathcal{U}_{safe}(t) = \mathcal{U}_{h_1} \cap \mathcal{U}_{h_2} \cap \mathcal{U}_{h_3}(t)$. As $C_1 \cap C_2 \cap C_3(t) \neq \emptyset$,  $\mathcal{U}_{safe}(t)$ is also non-empty when $u \in \mathbb{R}$. After formulating the CBF constraints, we synthesize the controller.

\begin{figure*}
\centering
\setkeys{Gin}{width=0.233\textwidth}
\subfloat[$\mbox{HC-I}$: $h_1(x)\geq 0$. The ego vehicle always maintains the safe distance from the lead vehicle. 
          \label{fig:h1(x) vs Time}]{\includegraphics{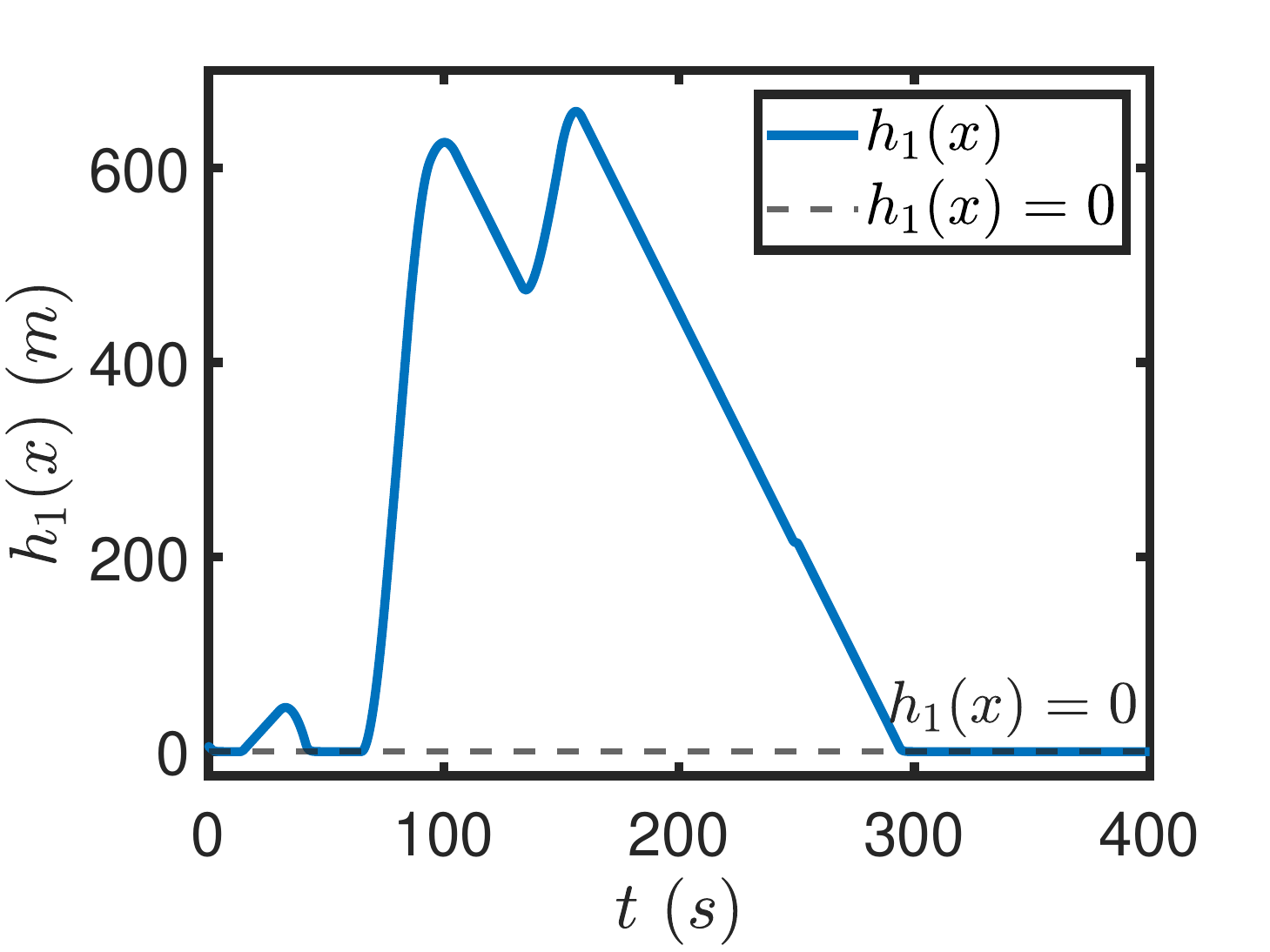}}
    \hfill
\subfloat[$\mbox{HC-II}$: $V_f\leq V_{max}$. The speed profile of the lead vehicle and the ego vehicle in $Km/h$ vs. time (s). 
         \label{fig:Vr_vs_Time}]{\includegraphics{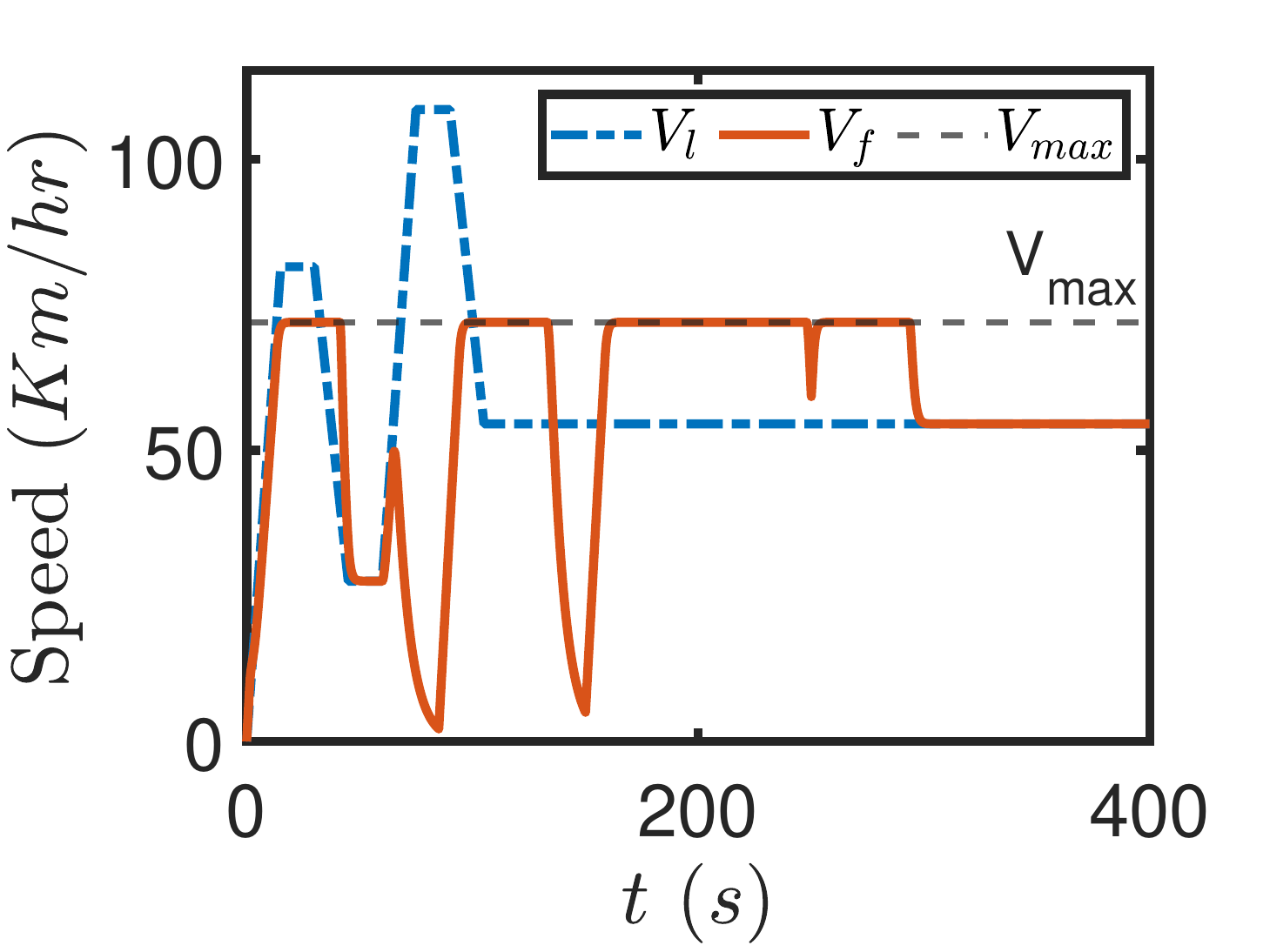}}
    \hfill
    \subfloat[ $\mbox{HC-III:} \bar{h}_3(t,\mathbf{x})\geq 0$. When $\bar{h}_3(t,\mathbf{x})\geq 0$ then the ego vehicle obeys the traffic signals. 
          \label{fig:h3 vs Time}]{\includegraphics{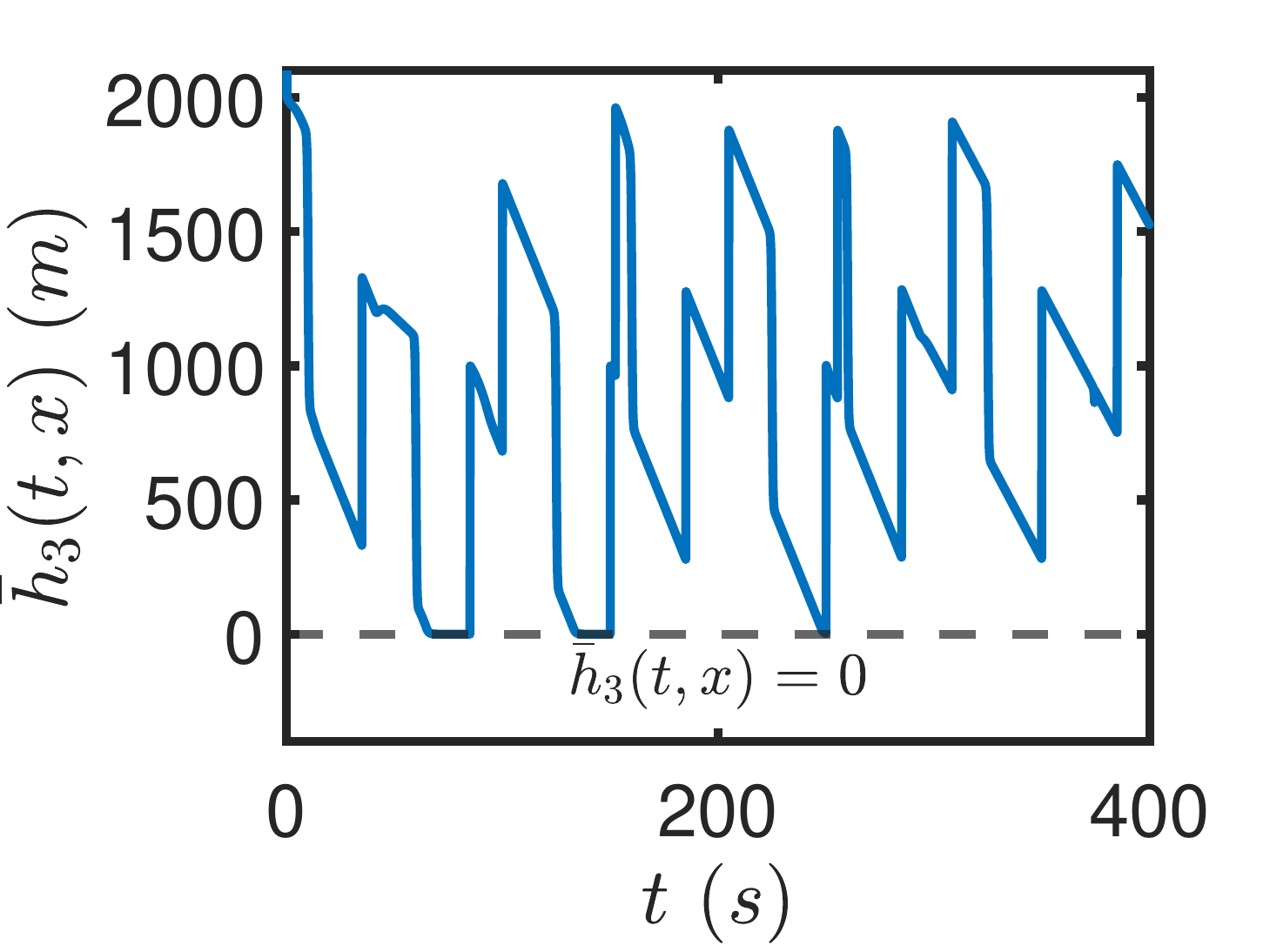}}
          \hfill
\subfloat[Wheel Force (N/mg) vs. time (s).
          \label{fig:u vs Time}]{\includegraphics{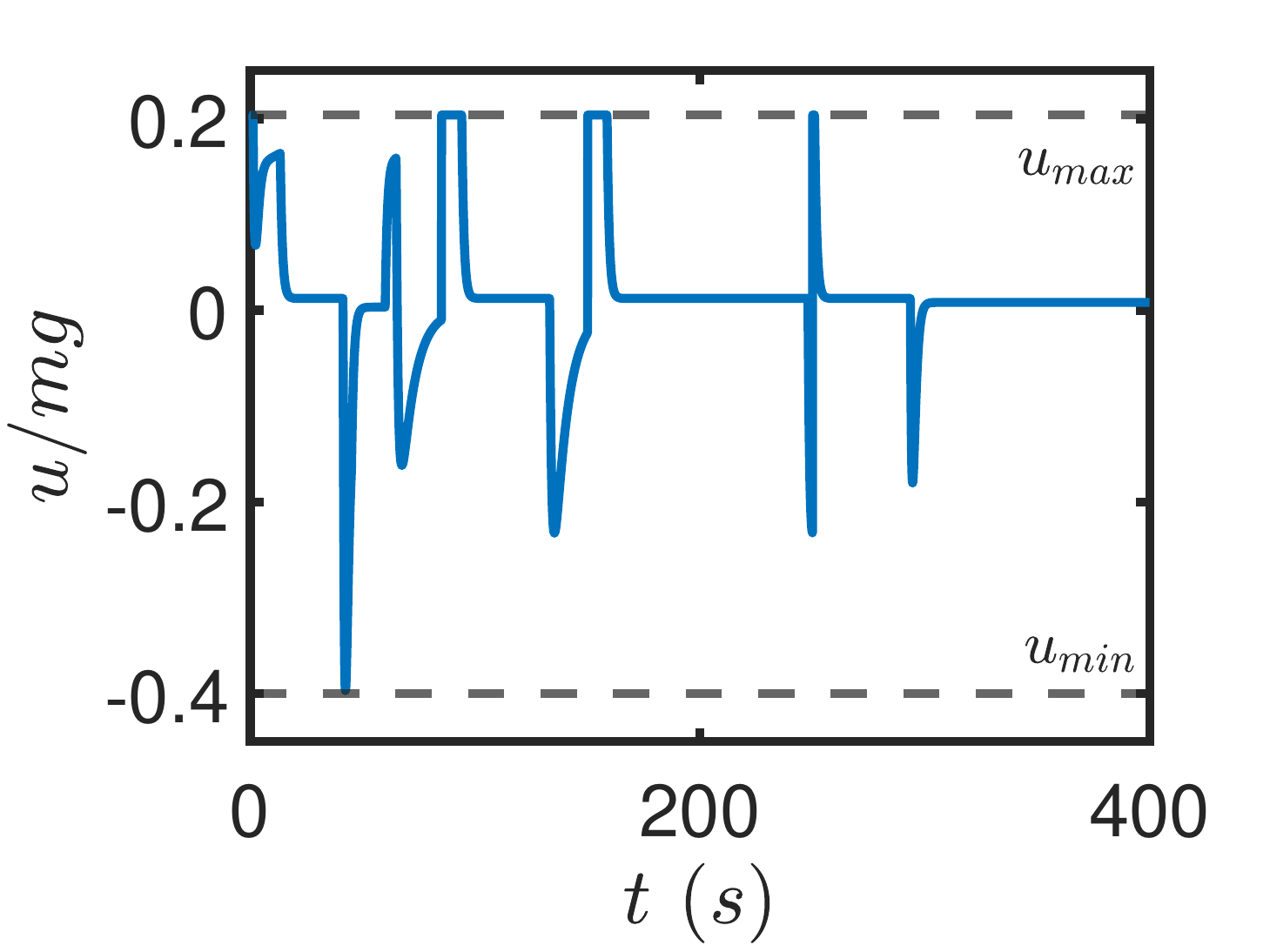}}
\caption{Simulation results.}
\label{fig:traffic_signals}
\end{figure*}

\subsection{Control Synthesis Using Quadratic Programming \label{Sec:Control QP}}

A nominal controller provides $u_{nom}$ to satisfy the soft constraints $V_f\rightarrow V_l$, $\delta \rightarrow 0$. If $u_{nom}\in \mathcal{U}_{safe}(t)$ then the input given by the nominal controller satisfies all the hard constraints. When $u_{nom} \notin \mathcal{U}_{safe}(t)$ then we solve the following QP problem~\cite{ames2019control}. 
 \begin{align}
 \label{eq:QP-CBF-Optimization}
 u &= \underset{u \in \mathcal{U}}{\mbox{argmin}}\; |u-u_{nom}|^2 \notag\\
 & \mbox{s.t.} \begin{cases}
 h\mu(u) \leq V_r +h_1(\mathbf{x})\\
\mu(u) \leq V_{safe}-V_f\\
\mu(u) \leq \frac{d^2h_{3it}(t)}{dt^2} + K_{\alpha}\begin{pmatrix} h_{3,i}(t,\mathbf{x}) \\  \frac{dh_{3it}}{dt} -V_f \end{pmatrix}\\
\mu = \frac{1}{m}(u-F_r)
\end{cases} 
\end{align}
As $\mathcal{U}_{safe}(t) \neq \emptyset$, the QP problem is always feasible when $u\in \mathcal{U}=\mathbb{R}$. In practice, the input force is constrained to $\mathcal{U}=[u_{min},u_{max}]$, which could make the QP problem infeasible. Therefore, we modify the CBF so that QP problem remains feasible under the input constraints. We also modify the CBF from relative degree $2$ to relative degree $1$ which is more intuitive and easier to implement.

\subsection{Input Constraints}

The input force is constrained, i.e., $\mathcal{U}= [-a_{min}m,a_{max}m]$ with $a_{min}>0$, $a_{max}>0$.  Therefore, we obtain $-a_{min}-\frac{F_r}{m}\leq \mu \leq a_{max}-\frac{F_r}{m}$. {Since $F_r\geq 0$, the minimum acceleration available is equal to $-a_{min}-\frac{Fr}{m}$, where the first term is the braking acceleration provided by the vehicle and $F_r/m$ is the extra braking acceleration provided by the aerodynamic and frictional forces. The stopping distance for the ego vehicle following a lead vehicle $V_l$ is given by $\frac{(V_l-V_f)^2}{2(a_{min}+F_r/m)}$. As a worst case scenario, we assume that all the braking force is provided by the vehicle. Therefore, after ignoring $\frac{F_r}{m}$, we can say that $\mu \geq -a_{min}$.} 
The CBF $h_1(t,\mathbf{x})$ can be easily modified by adding a stopping distance~\cite{ames2014control}:
\begin{align}
\begin{aligned}
& h_1(x) = X_r-hV_f-S_0-\frac{(V_l-V_f)^2}{2a_{min}},\\
& \frac{dh_1}{dt}(x) = V_r-h\mu+\frac{V_r}{a_{min}}\mu,\\
& \left(h-\frac{V_r}{a_{min}}\right)\mu \leq h_1(x)+V_r.
\end{aligned}
\end{align}
We also modify $h_{3i}(t,\mathbf{x})$ to incorporate input constraints. We define a headway from a stop line of the traffic signal. If the current velocity is $V_f$, then it will take a time  $\gamma=\frac{V_f}{a_{min}}$ for the ego vehicle to completely stop by applying its maximum braking effort. In the worst case scenario, $V_f = V_{max}$, $\gamma=\frac{V_{max}}{a_{min}}$. Therefore, the new CBF along with the CBF input constraints is given by
\begin{equation}
\label{eq:h3+Vf}
\begin{split}
\bar{h}_{3i}(t,\mathbf{x}) = h_{3it}(t) + p_i-X_f -\gamma V_f,\\
\gamma\mu \leq \bar{h}_{3i}(t,\mathbf{x})+\frac{dh_{3it}}{dt}(t)-V_f.
\end{split}
\end{equation}
This modified time-varying CBF 
has relative degree $1$ and enforces the input constraints.  

\subsection{Control Synthesis Using QP with Input Constraints}
The controller can be synthesised by solving the following QP problem with input constrains.
 \begin{align}
 \begin{split}
 \label{eq:QP-CBF-Optimization-input-constraints}
 u &= \underset{u \in \mathcal{U}}{\mbox{argmin}}\; |u-u_{nom}|^2\\
 & \mbox{s.t.} \begin{cases}
 \left(h-\frac{V_r}{a_{min}}\right)\mu(u) \leq h_1(x)+V_r\\
\mu(u) \leq V_{safe}-V_f\\
\gamma\mu (u) \leq \bar{h}_3(t,\mathbf{x})+\frac{dh_{3i}}{dt}(t)-V_f\\
-a_{min} \leq \mu(u) \leq a_{max}\\
\mu = \frac{1}{m}(u-F_r),
 \end{cases}\\
 \mathcal{U} &= [-a_{min}m,a_{max}m].
 \end{split}
 \end{align}
The above controller ensures that the system fulfills the safety and regulatory constraints. 

\section{Simulation Results \label{Sec:Simulation}}

We validate our approach on a road trip. A road scenario consists of a straight road with $6$ traffic signals, each $1$~Km away from the other. 
Traffic signals are not synchronized. Each traffic signal has its own timing sequence that is broadcast to the ego vehicle ahead of time. The values of the parameters used are $g=9.8\;\frac{m}{s^2}$, $m=1650~Kg$, $c_0=0.1\:N$, $c_1= 5\: \frac{N}{m/s}$, $c_2=0.25\frac{N}{m/s^2}$, $a_{max}=0.2\mbox{g} \frac{m}{s^2}$, $a_{min}=-0.4\mbox{g}\frac{m}{s^2}$, and $\tau=6$. The gains of the nominal controller are $k_1=7.12$, $k_2=3.24$, $k_3=0.4$. The ego vehicle and the lead vehicle are initially at rest and $X_r(0)=S_0=4.5$ holds with $X_f=0$, $X_l=4.5$. The simulation was performed in MATLAB. In Fig.~\ref{fig:Vr_vs_Time}, the lead vehicle is not following any traffic rules. It first speeds up, then maintains its velocity, and then decelerates. The lead vehicle violates the traffic signals and goes beyond the maximum speed limit of $V_{max}$. 

\begin{figure}[t]
         \includegraphics[width=\columnwidth]{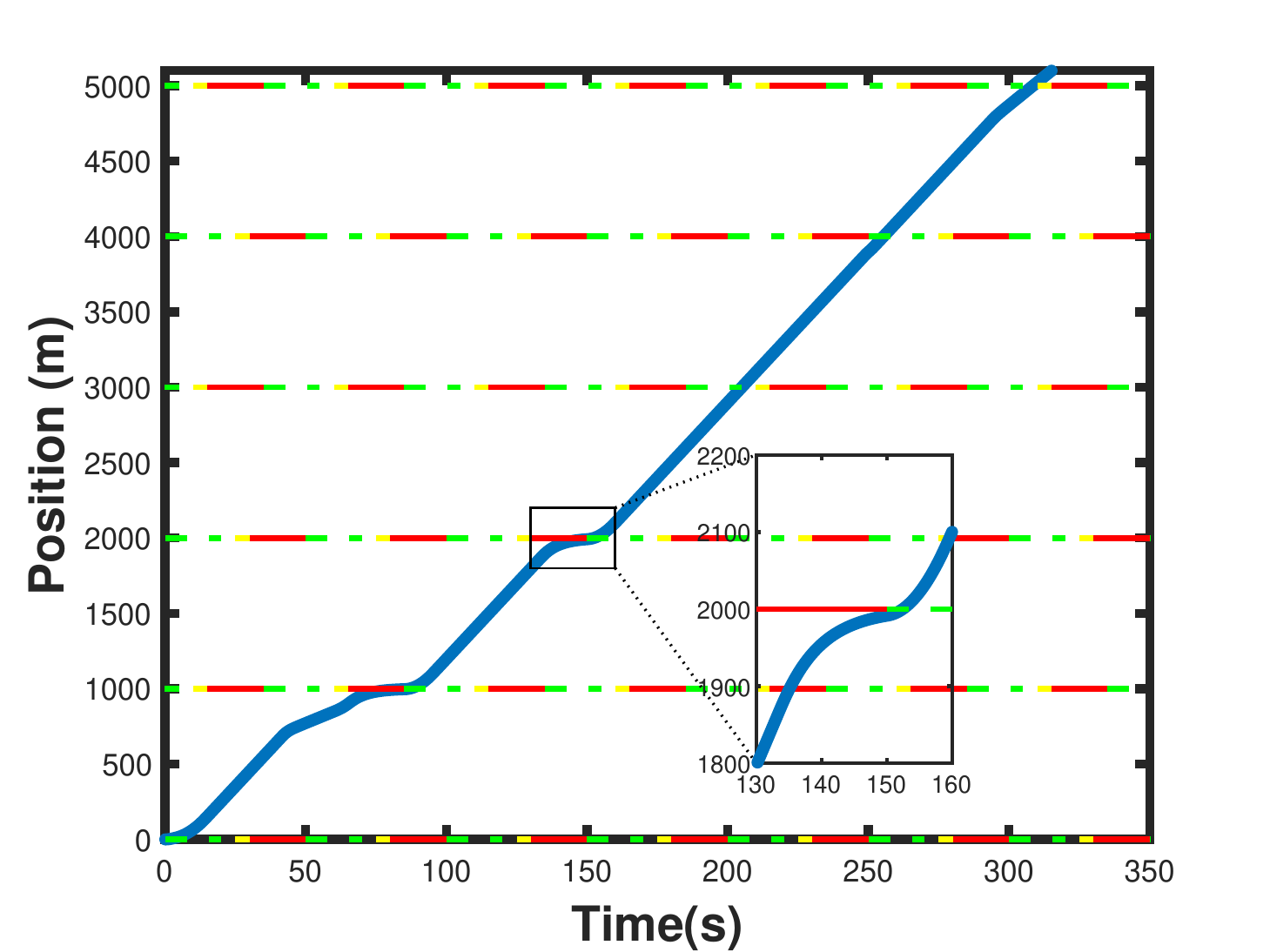}
         \caption{Position of the ego vehicle $X_f$ (m) vs. time (s). The horizontal lines mark the position of traffic signals while their color shows their state at a given time. The solid red lines show the intervals when the signals are red.}
         \label{fig:Xf vs Time}
\end{figure}

The ego vehicle follows the lead vehicle while observing all the hard constraints $\mbox{HC-I}$ (Fig.~\ref{fig:h1(x) vs Time}), $\mbox{HC-II}$ (Fig.~\ref{fig:Vr_vs_Time})  and $\mbox{HC-III}$ (Fig.~\ref{fig:h3 vs Time} and Fig.~\ref{fig:Xf vs Time}). Fig.~\ref{fig:h1(x) vs Time} shows that $\mbox{HC-I}:h_1(x)\geq 0$ is satisfied by the controller. Fig.~\ref{fig:Vr_vs_Time} shows that $\mbox{HC-II}:V_f\leq V_{max}$ is always satisfied and the ego vehicle keeps its speed below the maximum speed limit $V_{max}$. From $t=65\;s$ to $154\;s$, the lead vehicle violates the traffic signals and goes at a velocity higher than $V_{max}$ but the ego vehicle obeys the traffic signals while still keeping the velocity below $V_{max}$. $h_1(x)$ first increases from $t=65\;s$ to $t=154\;s$, then it decreases as the lead vehicle slows down but it is never negative. The wheel force in $N/mg$ is shown in Fig.~\ref{fig:u vs Time}. The input is constrained to the limit specified by the controller, i.e., $u_{min}\leq u \leq u_{max}$.

The controller always satisfies $\mbox{HC-III}$, i.e., the ego vehicle obeys the traffic signals. Fig.~\ref{fig:Xf vs Time} shows the trajectory of the ego vehicle with time. The height of the horizontal lines shows the position of the traffic signals, and the solid red horizontal lines show the interval where a traffic signal is red. The duration of green, red, and yellow signals are $25~s$, $20~s$, and $5~s$, respectively. The zoomed plot in the Fig.~\ref{fig:Xf vs Time} and the speed profile of the ego vehicle in Fig.~\ref{fig:Vr_vs_Time} show that the ego vehicle preemptively slows down before a red signal and smoothly accelerates when the signal is green, thereby avoiding unnecessary stops by applying comfortable braking force as set by the input constraints. 

The traffic rules are encoded by $\bar{h}_3(t,\mathbf{x})$, that is, the non-negativeness of $\bar{h}_3(t,\mathbf{x})$ means that the ego vehicle obeys the traffic rules. Fig.~\ref{fig:h3 vs Time} shows that $\bar{h}_3(t,\mathbf{x})\geq 0$ holds. The value of $\bar{h}_3(t,\mathbf{x})$ represents the safe distance from the next red signal. When $p_{i-1}<X_f\leq p_i$ and the $i^{th}$ signal is green,  $\bar{h}_3(t,\mathbf{x})$ represents the safe distance from the $(i+1)^{th}$ traffic signal at $p_{i+1}$. When $p_{i-1}<X_f\leq p_i$ and the $i^{th}$ traffic signal is yellow, then $\bar{h}_3(t,\mathbf{x})$  smoothly decreases. For example, initially, at $t=0$, the ego vehicle is at rest at $X_f=0$ and the traffic signal at $p_1=1~Km$ is green. Therefore, $\bar{h}_{3}(0,\mathbf{x(0)})\approx 2000$. When $p_1$ is yellow, then $\bar{h}_{3}(t,\mathbf{x})$ decreases. Later, when the upcoming traffic signal turns green, $\bar{h}_3(t,\mathbf{x})$ increases. When the upcoming traffic signal is yellow, then $\bar{h}_3(t,\mathbf{x})$ starts decreasing smoothly. The ego vehicle always follows the traffic signals and never crosses the stop line when the signal is red. 
The controller fulfills the soft constraint that $V_f \rightarrow V_l$ and $\delta=h_1(x)\rightarrow0$ as long as the hard constraints $\mbox{HC-I}$, $\mbox{HC-II}$ and $\mbox{HC-III}$ are not violated. 



\section{Conclusions}
\label{Sec:Conclusion}

We presented a correct-by-construction adaptive cruise control design method under safety and regulatory constraints with control barrier guarantees. The proposed regulated ACC obeys the traffic signals and speed limits while maintaining safe spacing from the lead vehicle. The rules for traffic signals are described in the form of piecewise $\mathcal{C}^m$ time-varying control barrier functions (TV-CBFs). 
We proved that, for a valid piecewise $\mathcal{C}^m$ TV-CBF, there exists a controller that renders the corresponding superlevel set forward-invariant. Given a valid piece-wise $\mathcal{C}^m$ TV-CBF, a switching-based controller can be synthesized using quadratic programming. 
Simulation results validate the efficacy of the proposed method. 


\section*{Acknowledgments}
This research was supported in part by the National Science Foundation (NSF) under Awards 1839842 and 1846524, the Office of Naval Research (ONR) under Award N00014-20-1-2258, the Defense Advanced Research Projects Agency (DARPA) under Award HR00112010003, and METRANS Transportation Center under the following grants: Pacific Southwest Region 9 University Transportation Center (USDOT/Caltrans) and the National Center for Sustainable Transportation (USDOT/Caltrans).






\bibliography{IEEEabrv.bib,IEEEexample.bib, references.bib}
\bibliographystyle{IEEEtran}


\end{document}